\documentclass[sigconf]{acmart}
\usepackage{multirow}
\usepackage{xcolor}
\usepackage{tabularray}
\usepackage{hyperref}
\usepackage{subcaption}

\usepackage{tabularx}

\usepackage{framed}
\usepackage{booktabs}
\usepackage{siunitx}
\usepackage{hyperref}

\definecolor{lightblue}{rgb}{0.8, 0.9, 1.0}

\newcommand{\pmark}[1]{\textsuperscript{\hypertarget{pmark:#1}{(\smash{#1})}}}

\usepackage[most]{tcolorbox}
\newtcolorbox{bluebox}[1][]{
  colback=blue!5!white,
  colframe=blue!75!black,
  boxsep=2pt,      % inner padding between frame and content
  top=2pt,         % space between frame and top content
  bottom=2pt,      % space between frame and bottom content
  left=5pt,        % space between frame and left content
  right=5pt,       % space between frame and right content
  width=\linewidth % set the width of the box to match the column width
}
\newtcolorbox{boxK}{
    colframe = cs-2!80!white,
    sharpish corners, % better drop shadow
    boxrule = 0pt,
    boxsep = 0pt,
    toptitle = 2pt,
    bottomtitle = 1pt,
    top = 4pt,
    bottom = 5pt,
    left = 5pt,
    right = 5pt,
    enhanced,
    fuzzy shadow = {0pt}{-1pt}{-0.5pt}{0.5pt}{black!20}, % {xshift}{yshift}{offset}{step}{options}
    fonttitle=\small\bfseries,
    fontupper=\small,
    title=Main Findings%Key Takeaways
}
\usepackage{blindtext}
\usepackage{balance}

\usepackage{listings}
\newtcblisting{promptbox}{
  listing engine=listings,
  listing only,
  colback=gray!4, colframe=gray!60!black,
  title={\texttt{CLAUDE.md}},
  left=1mm, right=1mm, top=1mm, bottom=1mm,
  breakable,
  listing options={style=prompt, basicstyle=\footnotesize\ttfamily} % <-- keep this LAST
}

\usepackage{mdframed}
\newmdenv[
  leftline=true,
  topline=false,
  bottomline=false,
  rightline=false,
  linecolor=blue,
  linewidth=2pt,
  backgroundcolor=white,
  skipabove=10pt,
  skipbelow=10pt,
  innerleftmargin=6pt,
  innertopmargin=4pt,
  innerbottommargin=4pt,
  nobreak=true
]{InterviewQuote}

\usepackage{fontawesome5} % for \faLightbulb

\balance

\AtBeginDocument{%
  }

\definecolor{main}{HTML}{CFCFCF}
\definecolor{sub}{HTML}{CFCFCF}
\definecolor{grayborder}{RGB}{180,180,180}
\definecolor{titlegray}{RGB}{230,230,230}
\definecolor{mygray}{gray}{0.9}

\newenvironment{boxC}{
    \MakeFramed{\hsize\linewidth\advance\hsize-\width\FrameRestore}\noindent%
}{%
    \endMakeFramed
}

\begin{document}

\title[Code for Machines, Not Just Humans]{Code for Machines, Not Just Humans: Quantifying AI-Friendliness with Code Health Metrics}

\author{Markus Borg}
\orcid{0000-0001-7879-4371}
\affiliation{%
  \institution{CodeScene and Lund University}
  \city{Malmö}
  \country{Sweden}
}
\email{markus.borg@codescene.com}

\author{Nadim Hagatulah}
\orcid{0009-0005-0840-1073}
\affiliation{%
  \institution{Lund University}
  \city{Lund}
  \country{Sweden}
}
\email{nadim.hagatulah@cs.lth.se}

\author{Adam Tornhill}
\orcid{XXX}
\affiliation{%
  \institution{CodeScene}
  \city{Malmö}
  \country{Sweden}
}
\email{adam.tornhill@codescene.com}

\author{Emma Söderberg}
\orcid{0000-0001-7966-4560}
\affiliation{%
  \institution{Lund University}
  \city{Lund}
  \country{Sweden}
}
\email{emma.soderberg@cs.lth.se}

\renewcommand{\shortauthors}{Borg et al.}

\begin{abstract}
We are entering a hybrid era in which human developers and AI coding agents work in the same codebases. While industry practice has long optimized code for human comprehension, it is increasingly important to ensure that LLMs with different capabilities can edit code reliably. In this study, we investigate the concept of ``AI-friendly code'' via LLM-based refactoring on a dataset of 5,000 Python files from competitive programming. We find a meaningful association between CodeHealth, a quality metric calibrated for human comprehension, and semantic preservation after AI refactoring. Our findings confirm that human-friendly code is also more compatible with AI tooling. These results suggest that organizations can use CodeHealth to guide where AI interventions are lower risk and where additional human oversight is warranted. Investing in maintainability not only helps humans; it also prepares for large-scale AI adoption.
\end{abstract}




\maketitle

\section{Introduction} \label{sec:intro}
For decades, the maxim has been that \emph{``programs must be written for people to read, and only incidentally for machines to execute''}~\cite{abelson_structure_1984}. Human-readable code is essential for maintaining secure, reliable, and efficient software development~\cite{sadowski_modern_2018,munoz_baron_empirical_2020,borstler_developers_2023}. But with the advent of AI-assisted coding, source code now has a broader audience: machines need to understand its intent, too.

Early field observations collected by Thoughtworks\footnote{\url{https://www.thoughtworks.com/}} suggest that AI-assisted coding tools perform better on high-quality code. In particular, well-factored modular code seems to reduce the risk of hallucination and lead to more accurate suggestions. They refer to this as \emph{``AI-friendly code design''} in their April 2025 Technology Radar, and discuss how established best practices so far align with AI-friendliness~\cite{thoughtworks_technology_2025}. If this observation holds, it implies that code optimized for human comprehension is also easier for Large Language Models (LLMs) to process and evolve.

The implications of AI-friendliness are profound: as of 2025, about 80\% of developers already use AI tools in their work~\cite{stack_overflow_2025_2025}, with adoption projected to grow rapidly. Gartner predicts an increase from 14\% in early 2024 to 90\% by 2028~\cite{walsh_magic_2025}. At the same time, less than 10\% of organizations have been reported to methodically track technical debt~\cite{martini_technical_2018}, and developers have been found to waste up to 42\% of their time due to poor code quality~\cite{stripe_developer_2018}. These figures suggest that much of today's production code may be structurally unfit for reliable AI intervention, increasing the risk of bugs and expensive rework.

In this paper, we investigate the relationship between code quality and AI-friendliness. We measure quality using the \textit{CodeHealth} (CH) metric, which has been validated as predictive of defects and development effort in previous studies~\cite{tornhill_code_2022,borg_increasing_2024,borg_ghost_2024}. The metric has also been used in recent industry-facing AI studies~\cite{borg_echoes_2025,meaden_compass_2025}. We use the success rate of AI-generated refactoring, i.e., improving the design of existing code without changing its behavior, as a proxy for AI-friendliness. Refactoring lets us use passing unit tests as an oracle for functional correctness. Thus, an AI refactoring is \textit{correct} if tests pass and \textit{beneficial} if CH increases. %We study two settings: zero-shot prompts for refactoring with five LLMs and agentic refactoring with Claude Code.

Our results confirm that human-friendly code is more compatible with AI tooling. LLMs tasked with refactoring have significantly lower break rates on code in the Healthy CodeHealth range ($CH\ge9$), with corresponding risk reductions of 15-30\%. Furthermore, we show that CH outperforms \textit{perplexity} (PPL, an LLM-intrinsic confidence metric) and Source Lines of Code (SLOC) as a predictor of refactoring correctness. 

These findings can support software organizations in the adoption of AI-assisted coding. We propose using CH to identify parts of the code that are ready for AI processing, as well as highlighting code with too much risk of breaking. More broadly, this research adds a missing piece to AI adoption: a shared code-quality metric that aligns humans and machines. While earlier research positioned code quality as a business imperative, we posit that code quality is a prerequisite for safe and effective use of AI -- which might prove existential for software organizations in the next decade.

The remainder of this paper is organized as follows. Section~\ref{sec:bg} describes background and related work. In Section~\ref{sec:method}, we introduce our research questions, dataset, and method. Section~\ref{sec:res} reports our results. Finally, Section~\ref{sec:disc} discusses the implications before Section~\ref{sec:conc} concludes with directions for future work.

\section{Background and Related Work} \label{sec:bg}
This section first introduces CH and PPL, then presents related work on code comprehension and AI refactoring.

\subsection{Maintainability and CodeHealth}
CodeHealth™ (CH) is a quality metric used in the CodeScene software engineering intelligence platform. Its goal is to capture how cognitively difficult it is for human developers to comprehend code. CodeScene identifies \textit{code smells}~\cite{lacerda_code_2020}, e.g., God classes, deeply nested logic, and duplicated code. For Python, which we target in this paper, CodeScene detects 25 code smells. 

CodeScene combines the number and severity of detected smells into a file-level score from 1 to 10. Lower scores indicate higher cognitive load for humans, i.e., higher maintenance effort. CodeScene categorizes files as belonging to one of three CH intervals: \textit{Healthy} (CH $\ge 9$), Warning ($4 \le \text{CH} < 9$), and Alert (\text{CH} $< 4$). In this study, we refer to both Warning and Alert files as \textit{Unhealthy}.

We have previously validated the CH metric in a series of studies. In a study on the manually annotated \textit{Maintainability Dataset}~\cite{schnappinger_defining_2020}, we reported that CH aligns better with human maintainability judgments than competing metrics and the average human expert~\cite{borg_ghost_2024}. Furthermore, we have validated the metric from a business perspective through the association between CH on the one hand and file-level defect density and development time on the other hand~\cite{tornhill_code_2022,borg_increasing_2024}. Building on our previous work, we now investigate the relation between CH and AI\nobreakdash-friendliness.

\subsection{LLM Confidence and Perplexity} \label{sec:ppl}
Several metrics have been proposed to assess the confidence of LLM output. Internal metrics such as entropy, mutual information, and PPL can potentially be used as proxies of output correctness and hallucination risk. In the software domain, Sharma \textit{et al.} showed that higher entropy and mutual information correlate with lower functional correctness of the generated code~\cite{sharma_assessing_2025}. Earlier work by Ray \textit{et al.} also showed that code containing defects had higher entropy than correct code~\cite{ray_naturalness_2016}, which in turn builds on the foundational work on the naturalness of code by Hindle \textit{et al.}~\cite{hindle_naturalness_2016}. 

Ppl, originally proposed for speech recognition tasks~\cite{jelinek_perplexity_1977}, is the exponential of the prediction model's average entropy over a sequence. Considering entropy to represent ``average surprise,'' PPL rather turns this into a scale of ``how many choices.'' It is used as a measure of how confident an LLM is in predicting the next token based on the previous tokens -- a higher score means lower confidence. Mathematically, let $t_{1:N}$ be a token sequence.
The average cross-entropy ($H$) is:
\[
H = -\frac{1}{N}\sum_{i=1}^{N}\log p(t_i \mid t_{<i}).
\]
PPL is $\exp(H)$. In this work, we get PPL directly from the five Hugging Face models under study~\cite{hugging_face_perplexity_2025}.

\subsection{Code Comprehension and Perplexity}
Gopstein \textit{et al.} coined the term \textit{atoms of confusion} to refer to small, isolated patterns in code that confuse humans~\cite{gopstein_understanding_2017}. An atom is the smallest unit of code that can cause confusion. Through controlled experiments, they identified a set of atoms that significantly increased the rate of misunderstandings. Examples include ``Assignment as Value,'' e.g., \texttt{V1 = V2 = 3;} and ``Logic as Control Flow,'' e.g., \texttt{V1 \&\& F2();}.

Building on the atoms concept, Abdelsalam \textit{et al.} discussed the risks of comprehension differences between human programmers and LLMs in the current hybrid development landscape~\cite{abdelsalam_how_2025}. They argue that if humans and LLMs are confused by different characteristics in code, there is a risk of misalignment. By comparing LLM PPL and EEG responses, they found that LLMs and humans struggle with similar issues in code. A previous controlled experiment by Casulnuovo \textit{et al.}, also suggests a connection between human comprehension and how surprising a code snippet is to LLMs~\cite{casalnuovo_does_2020}.

Kotti \textit{et al.} explored the PPL of different programming languages \cite{kotti_fools_2025}. Based on 1,008 files from GitHub projects representing different languages, they found systematic differences between strongly-typed and dynamically typed languages. For example, across LLMs, they found that Perl resulted in high PPL and Java in low. They speculate that their PPL findings could be used to assess the suitability of LLM-based code completion in specific projects. In this study, we explore the PPL concept for Healthy vs. Unhealthy Python code.

\subsection{AI Refactoring}
Research in automated code refactoring has, like so many software engineering applications, been disrupted by LLMs. Several studies investigate the potential of various combinations of LLMs and prompts to improve design and remove code smells effectively. Pomian \textit{et al.} introduced the tool EM-Assist~\cite{pomian_next-generation_2024} to automatically suggest and perform extract method refactorings. The same team has continued working on the more sophisticated refactoring operation \textit{move method}~\cite{batole_leveraging_2025}. Tornhill \textit{et al.} have developed the tool ACE to automatically remove five CodeScene code smells~\cite{tornhill_ace_2025}.

During 2025, agentic AI has been a major trend in industry and research. Coding agents are driven by LLMs, but they typically claim to 1) understand codebases beyond limited context windows and to 2) maintain a memory. This enables agents to take on larger tasks and operate more autonomously. The arguably most popular agent in industry at this time of writing is Anthropic's Claude Code, carefully described by Watanabe \textit{et al.}~\cite{watanabe_use_2025}. We refer to reviews by He \textit{et al.}~\cite{he_llm-based_2025} and Wang \textit{et al.}~\cite{wang_agents_2025} for contemporary overviews of the academic literature.

Refactoring is one of the many tasks investigated for coding agents. For example, Xu \textit{et al.} presented MANTRA, a multi-agent framework for refactoring~\cite{xu_mantra_2025}. MANTRA organizes three agents into 1) \textit{developer}, 2) \textit{reviewer}, and 3) \textit{repair} roles for the refactoring task, and outperforms direct LLM-usage. However, Claude is also a capable refactoring agent despite not being a multi-agent solution. In this study, we study Claude (v2.0.13) as a representative example of contemporary agentic AI refactoring. Moreover, MANTRA is not publicly available at the time of this writing.

\section{Method} \label{sec:method}
Our goal is to explore how code characteristics influence the capabilities of AI refactoring, with a particular focus on CH. We formulate three Research Questions (RQs) that explore code characteristics in light of AI-friendliness. First, we take an LLM-intrinsic perspective and study PPL. This connects to related work and provides a baseline. Second, we look at how the CH is associated with the success rate of a downstream AI task. Third, we examine the predictive power of CH compared to SLOC and PPL.

\begin{itemize}
    \item[\textbf{RQ$_1$}] How does perplexity differ between Healthy and Unhealthy code?
    \item[\textbf{RQ$_2$}] How does the AI refactoring break rate differ between Healthy and Unhealthy code?
    \item[\textbf{RQ$_3$}] To what extent can CodeHealth predict the AI refactoring break rate?
\end{itemize}

%\begin{boxC}
% { intrinsic: Does high codehealth signal low perplexity?}
%\textbf{RQ$_1$}. How does code perplexity differ between healthy and unhealthy code?
%\end{boxC}

%As a first step, we explore the foundations from an information theory perspective using the internal LLM metric perplexity. This is motivated by previous work suggesting that humans and machines are confused by similar constructs~\cite{abdelsalam_how_2025}. We build on Kotta et al.’s work on comparing perplexity across programming languages~\cite{kotti_fools_2025}, but instead focus on code with different CodeHealth.

%\begin{boxC}
% { downstream: Is codehealth associated with AI refactoring success?}
%\textbf{RQ$_2$}. How does the AI refactoring break rate differ between healthy and unhealthy code?
%\end{boxC}

%We measure both passing test cases and CodeHealth improvements. We do this for a series of LLMs of various size using zero-shot prompts. Second, we try a state-of-the-art coding agent.

%\begin{boxC}
% { predictive: Can CodeHealth serve as a practical predictor of AI refactoring outcomes?}
%\textbf{RQ$_3$}. To what extent can CodeHealth predict the success rate of AI refactoring?
%\end{boxC}

%Train a decision tree maybe? Or logistic regression with co-variates: CodeHealth, token count, anything else?

%RQ1: How high is the success rate for AI-refactorings in general?
%RQ2: To what extent can CodeHealth predict AI success rate? (TOTHINK: it's more than a correlation here, isn't it?
%RQ3: What is the recommended level of CodeHealth for AI-friendly code?

\subsection{Dataset Creation}
We sample from the CodeContests dataset hosted on GitHub\footnote{\url{https://github.com/google-deepmind/code_contests}} by Google DeepMind. The dataset was introduced by Li \textit{et al.}~\cite{li_competition-level_2022} as training data for AlphaCode and contains more than 12 million solutions (correct and incorrect) to competitive programming problems from five sources. We study code in this domain because the problems come with carefully crafted test cases that verify functional correctness, providing a practical oracle after refactoring.

We construct a dataset of 5,000 solutions based on four design choices. First, we decided to focus on solutions written in Python to increase novelty and convenience. While CodeContests also contains solutions in Java and C++, Java has been extensively studied in refactoring research, and C++ has a more complex compile-and-test procedure. Second, we require at least one CodeScene code smell in the solutions. Removing code smells is a realistic refactoring goal. Third, we chose to only study solutions containing between 60 and 120 SLoC. There is a strong correlation between maintainability and size~\cite{sjoberg_quantifying_2013}, thus we control for this, at least partly, already during the sampling. Fourth, we actively seek diversity in the dataset, as many solutions are highly similar. We use CodeBleu~\cite{ren_codebleu_2020} for similarity calculations using the following weights: \texttt{n-gram=0.1}, \texttt{weighted-n-gram=0.4}, \texttt{ast-match=0.5}, and \texttt{dataflow-match=0}.

The practical sampling process followed these steps (number of remaining solutions in parentheses):
\begin{enumerate}
\item Download the full CodeContests dataset (13,210,440).
\item Filter to only Python~3 solutions (1,502,532).
\item Remove all identical (Type~1 clone) solutions (1,390,531).
\item Remove all solutions with no CodeScene code smells (88,156).
\item Strip comments and unbound string literals.% (using Tree-sitter~\cite{brunsfeld_and_contributors_tree-sitter_2018}).
\item Filter to solutions with $60$–$120$ SLOC (18,074).
\item Partition into two strata: Healthy and Unhealthy code\footnote{A minor threshold-specification issue in the sampling script treated CH $=9$ as Unhealthy. All analyses in this paper use the intended convention (Healthy: $\ge 9$).}.%(CH $>9$) and Unhealthy code (CH $\leq9$)
\item For each stratum, repeat until 2,500 solutions are included:
\begin{enumerate}
    \item Sample a random candidate solution.
    \item Run its corresponding test cases; skip if any test case fails.
    \item Compute CodeBleu similarity to existing samples; skip if similarity $\geq 0.9$.
    \item Add the solution to the stratum.
\end{enumerate}
\item Merge the two strata to constitute the final dataset (5,000).
\end{enumerate}

%Section~\ref{sec:res-desc} presents descriptive statistics of the dataset.

\subsection{Selection of Large Language Models}
We select six LLMs for evaluation. Five are open-weight models with about 20-30B parameters, runnable on our local datacenter. We select four models based on popularity and download statistics from Hugging Face and complement them with an LLM recently published by IBM -- we refer to these as \textit{medium-sized LLMs}. Moreover, we include a State-of-the-Art (SotA) LLM that we prompt using Anthropic's API. 

\begin{itemize}
    \item[\textbf{Gemma}] gemma-3-27b-it (Google) -- Mar 2025.
    \item[\textbf{GLM}] GLM-4-32B-0414 (Zhipu~AI) -- Apr 2025.
    \item[\textbf{Granite}] Granite-4.0-H-Small (IBM) -- Oct 2025.
    \item[\textbf{GPT}] gpt-oss-20b (OpenAI) -- Aug 2025.
    \item[\textbf{Qwen}] Qwen3-Coder-30B-A3B-Instruct (Alibaba Cloud) -- Aug 2025.
    \item[\textbf{Sonnet}] claude-sonnet-4-5-20250929 (Anthropic) -- Sep 2025.
\end{itemize}

We set the sampling temperature to 0.7 for all LLMs to enable refactoring diversity under a uniform setting and we cap generation to a maximum of 8,192 new tokens. All other settings use defaults.

\subsection{RQ$_1$ Perplexity}
We extract PPL scores for all 5,000 samples for the five LLMs under study. Then we split the samples into Healthy and Unhealthy and state the following null hypothesis for each LLM:
\begin{itemize}
    \item[$H_{0}$:] The perplexity distributions for Healthy and Unhealthy code are identical.
\end{itemize}

Descriptive statistics revealed a small set of outliers with extremely high PPL. Manual analysis showed specific patterns that substantially inflate the LLMs' next-token uncertainty. In competitive programming, some contestants hard-code astronomically big integers or very long strings for later use in the solutions. To mitigate this, we applied one-sided robust z-score filtering and removed samples with z>2.5 from the output from the LLMs (about 5\%).

PPL scores are typically right-skewed, and we assess normality using the Shapiro–Wilk test. Even after filtering upper-tail outliers, we rejected normality for all LLM outputs. Subsequently, we rely on non-parametric Mann-Whitney~U tests for significance testing. Finally, we use Holm correction (Holm-Bonferroni step-down) to adjust for multiple comparisons across the five models (family-wise $\alpha$=0.05). We report two-sided Holm-corrected p-values and Cliff’s $\delta$ as the effect size.

\subsection{RQ$_2$ Refactoring Break Rate} \label{sec:method-rq2}
We tasked all medium-sized LLMs to refactor the 5,000 samples. For Sonnet, which has a high token price, we refactor 1,000 random samples equally split between Healthy and Unhealthy. For each combination of LLM and sample, we used the same prompt following a general structure, designed to be generic:
\begin{itemize}
    \item[(A)] Context in the form of a role description to steer the model toward the right parts of its knowledge.
    \item[(B)] A concrete task for the model to perform.
    \item[(C)] Instructions for how the model should format the response.
    \item[(D)] Input data for which the task shall be completed.
\end{itemize}

\begin{tcolorbox}[colback=gray!5,colframe=gray!60!black,title={Refactoring prompt},
                  left=1mm,right=1mm,top=1mm,bottom=1mm]
\ttfamily
\pmark{A}Act as an expert software engineer.\pmark{B}Your task is to refactor the following Python code for maintainability and clean code.\pmark{C} Respond ONLY with the complete, refactored Python code block. Do not add any explanations, comments, or introductory sentences.\\\pmark{D}Original code to refactor: \texttt{\textasciigrave\textasciigrave\textasciigrave\ python\ \textless CODE\textgreater\ \textasciigrave\textasciigrave\textasciigrave}
\end{tcolorbox}

We complemented the LLMs with Claude (v2.0.13) to also investigate a SotA agentic approach to refactoring. We selected Claude specifically because it is considered industry-leading and currently tops public SWE-bench results~\cite{jimenez_swe-bench_2024}. While Claude operates in an interactive mode in the terminal, we controlled the environment and presented the task in a manner consistent with the LLM setup. Claude is a costly service, and we decided to target the same 1,000 samples as for Sonnet. We aimed at a final cost of less than \$100 for Anthropic's solutions.

We organized batches of random samples in separate folders to mimic how a developer might work, starting with a small sample of 20 files for a pilot run -- followed by 200, 400, and 380. We provided the instructions (A--C) in a CLAUDE.md file in the common parent folder, appended with: ``\texttt{Each file in this folder is independent; you do not have to worry about dependencies between them.}''

For each sample folder used with Claude, we added a configuration file (\texttt{.claude/settings.json}) to
pin the model version and restrict the agent's tool use. Specifically, we set the model to
\texttt{claude-sonnet-4-5-20250929}, explicitly disabled \texttt{Bash}, \texttt{WebFetch}, and \texttt{WebSearch}, and also disabled any MCP use.

For each refactoring session, we followed these steps:
\begin{enumerate}
    \item Launch Claude in the current folder.
    \item Answer yes to ``Do you trust the files in this folder?''
    \item Ask Claude ``Can you see CLAUDE.md in the parent folder?'' and grant read permission.
    \item Give the instruction: ``Refactor the Python files in this folder for maintainability.''
    \item When Claude requests edit permission, approve editing of all files for the session.
\end{enumerate}

After each single refactoring pass, we recorded descriptive statistics of the output code, calculated its CH, and executed the corresponding test cases to determine whether behavior was preserved. If any test cases failed, we refer to the refactoring as \textit{broken}. We state the following null hypothesis for each refactoring approach:
\begin{itemize}
  \item[$H_{0}$:] There is no difference in refactoring break rate between Healthy and Unhealthy code.
\end{itemize}

We compare break rates using a chi-square ($\chi^2$) test of independence and report Risk Difference (RD) and Relative Risk (RR) with 95\% confidence intervals. We control family-wise error ($\alpha$=0.05) using Holm correction, in line with RQ$_1$.

\subsection{RQ$_3$ Predictive Power of CodeHealth}
We train decision trees on the data from the medium-sized LLMs' refactoring outputs to investigate which code features influence the probability that a refactoring breaks its test suite. Our feature set aims to cover three complementary dimensions while avoiding redundancy, motivated as follows:
\begin{itemize}
    \item \textbf{CodeHealth} is our primary explanatory variable of interest, reflecting human-oriented maintainability. 
    \item \textbf{Perplexity} is included because the LLMs' internal confidence can carry a predictive signal. Moreover, RQ$_1$ found that it is orthogonal to CH (see Section~\ref{sec:res-rq1}).       
    \item \textbf{SLOC} is included as a simple size metric, which often is an effective proxy for structural complexity. Kotti \textit{et al.}~\cite{kotti_fools_2025} (and our corroborating results for RQ$_1$) found that SLOC and PPL are not correlated.
    \item \textbf{Token count} is \emph{excluded} to minimize redundancy as it is correlated with SLOC (and showed small but consistent correlations with PPL in RQ$_1$).
\end{itemize}

We use a fixed, shallow decision tree (max depth = 3, min samples per leaf = 25, class-weighted) to prioritize interpretability and comparability across results for the six LLMs. We perform 5-fold cross-validation with these fixed hyperparameters, then refit the tree on all data for final visualization and rule extraction. Finally, we report the area under the ROC curve (AUC) for the final decision trees, i.e., a threshold-independent metric robust to class imbalance. Note, however, that we train decision trees for explanatory purposes rather than accurate predictions.

As a robustness check, we fit logistic regression models on the same data and features. Logistic regression provides Odds Ratios (OR), which show the change in odds that tests pass associated with a one-standard-deviation increase in a predicting feature, while holding all others constant.

\section{Results} \label{sec:res}
This section first presents descriptive statistics of the dataset, followed by results for the three RQs. All related Jupyter Notebooks are available in the replication package~\cite{borg_ai-friendliness_2026}. 

\subsection{Descriptive Statistics} \label{sec:res-desc}
\begin{figure}[htbp]
  \centering
  % --- Left subplot ---
  \begin{subfigure}{0.495\linewidth}
    \centering
    \includegraphics[width=\linewidth]{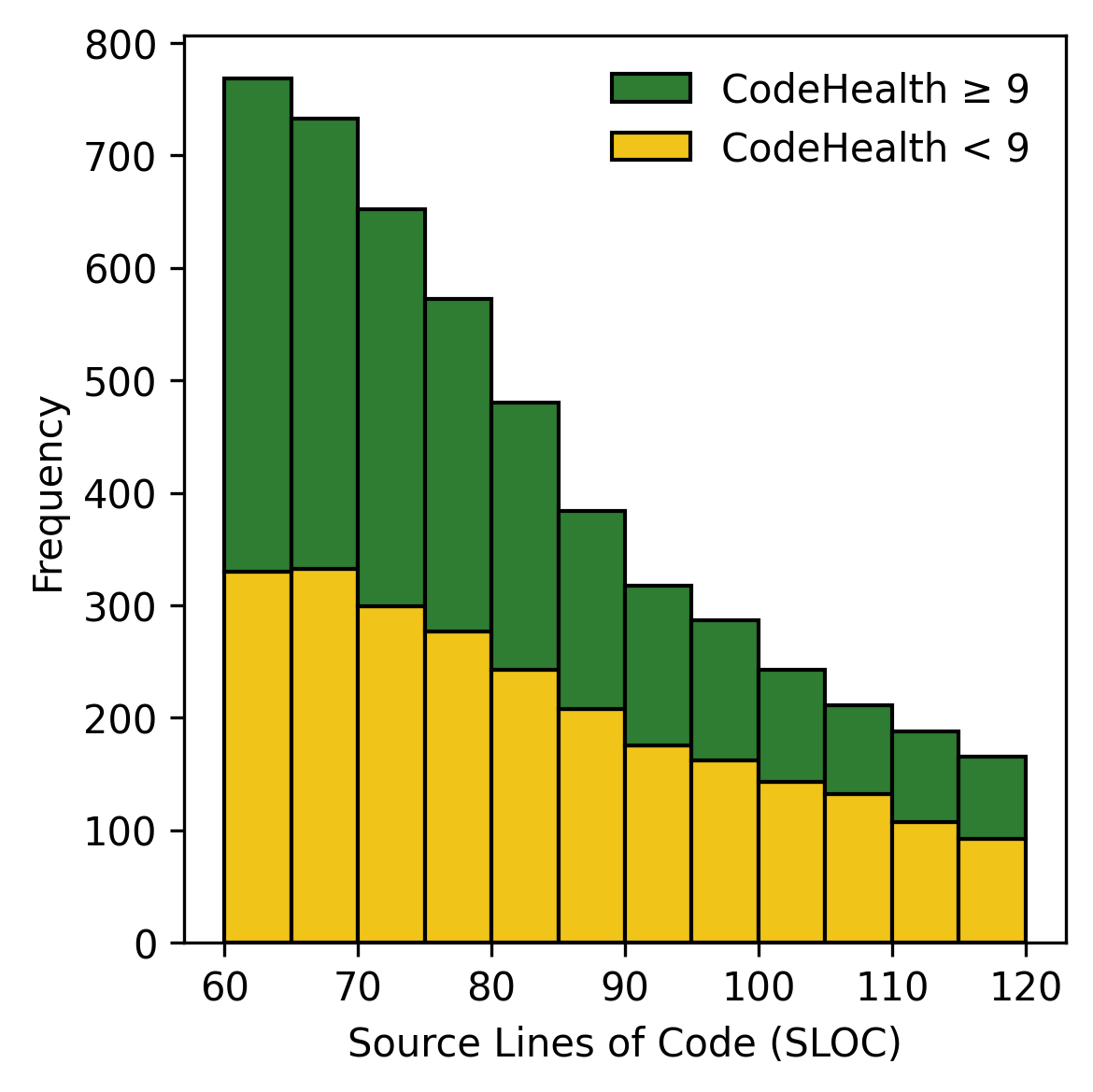}
    \caption{SLOC distribution.}
    \label{fig:hist_sloc}
  \end{subfigure}
  \hfill
  % --- Right subplot ---
  \begin{subfigure}{0.495\linewidth}
    \centering
    \includegraphics[width=\linewidth]{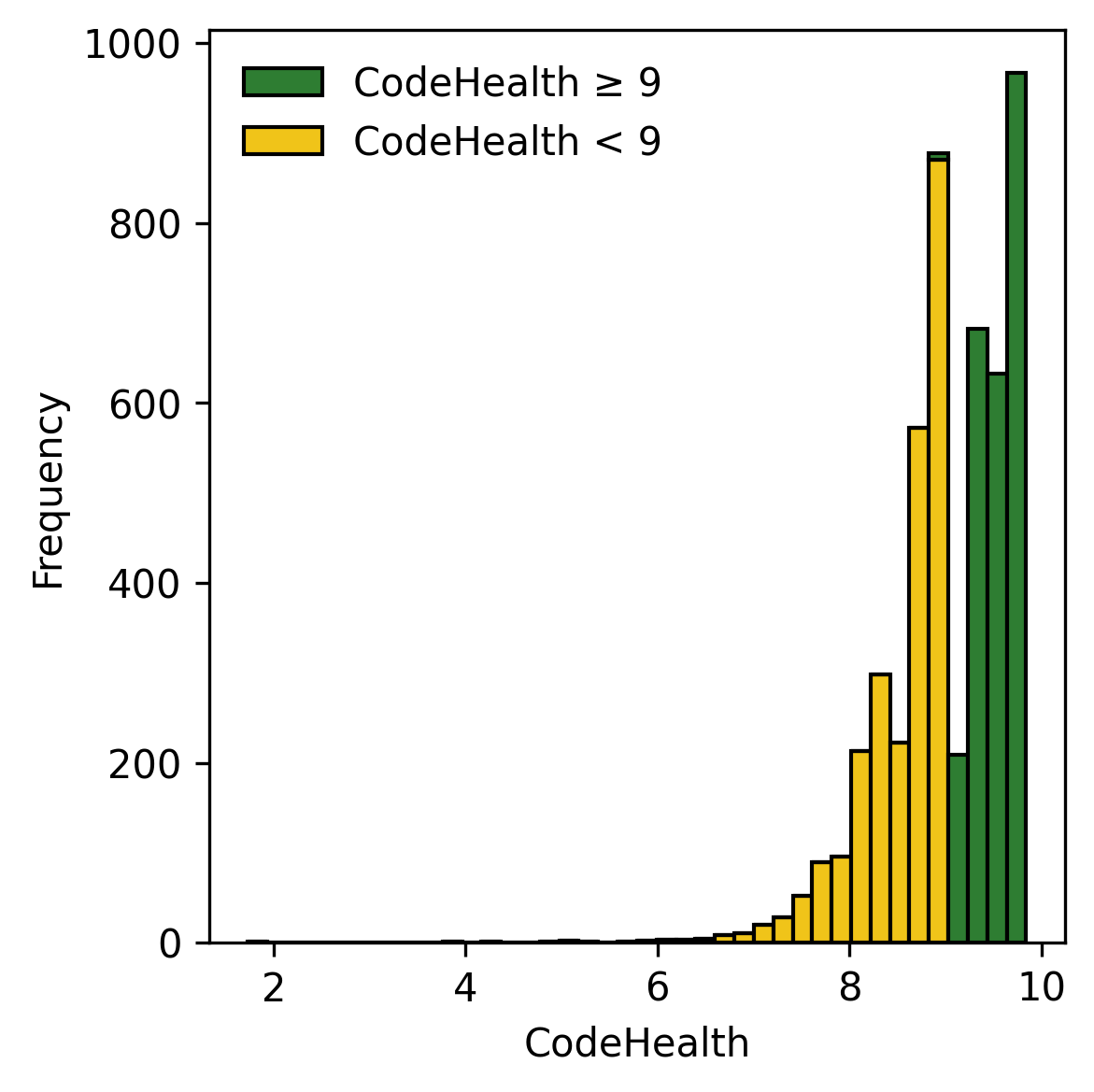}
    \caption{CodeHealth distribution.}
    \label{fig:hist_codehealth}
  \end{subfigure}
  \caption{Distributions of SLOC and CodeHealth.}
  \label{fig:metric_distributions}
\end{figure}

Figure~\ref{fig:metric_distributions} shows the SLOC and CH distributions of the 5,000 solutions in the dataset. For SLOC, the solutions are concentrated around the lower bound and then declines steadily with size. For CH, the distribution is left-skewed and we observe the effect of the stratified sampling. For both strata, the density piles up for solutions with higher CH, which explains the dip just over CH $=9$. Overall, the CH distributions resemble patterns reported in previous work~\cite{tornhill_code_2022,borg_increasing_2024}, though the skew is less pronounced for the short competitive programming tasks under study in this paper. 

CodeScene identifies nine different code smells in the dataset. The five most common smells (and their counts) are: 1)
\textbf{Bumpy Road Ahead} (n=4,901). A function contains multiple chunks of nested control structures, suggesting missing abstractions that make the code harder to understand. 2) \textbf{Complex Method} (n=3,572). A function has a high cyclomatic complexity, meaning it contains many independent logical execution paths. 3) \textbf{Deep, Nested Complexity}] (n=2,433). Control structures, such as loops and conditionals, are nested within each other to multiple levels. 4) \textbf{Complex Conditionals} (n=1,328). There are expressions that combine multiple logical operations, such as conjunctions and disjunctions, within a single condition. 5) \textbf{Excessive Function Arguments} (n=724). A function take too many parameters, indicating that the function is doing too much or that it lacks a proper abstraction.

%CodeScene identifies nine different code smells in the dataset. The five most common smells (and their counts) are:
%\begin{itemize}
%    \item \textbf{Bumpy Road Ahead} (n=4,901). A function contains multiple chunks of nested control structures, suggesting missing abstractions that make the code harder to understand.
%    \item \textbf{Complex Method} (n=3,572). A function has a high cyclomatic complexity, meaning it contains many independent logical execution paths.
%    \item \textbf{Deep, Nested Complexity}] (n=2,433). Control structures, such as loops and conditionals, are nested within each other to multiple levels.
%    \item \textbf{Complex Conditionals} (n=1,328). There are expressions that combine multiple logical operations, such as conjunctions and disjunctions, within a single condition.    
%    \item \textbf{Excessive Function Arguments} (n=724). A function take too many parameters, indicating that the function is doing too much or that it lacks a proper abstraction.
%\end{itemize}

\subsection{RQ$_1$ Perplexity} \label{sec:res-rq1}
Table~\ref{tab:rq1} summarizes PPL by the CH groups for each medium-sized LLM. We find that the PPL distributions are significantly different for all of them except Granite. However, the directions are mixed across models and the effect sizes are negligible for all models. Overall, our study yields a negative result, i.e., we find no practically meaningful association between PPL and CH.

\begin{table*}[t]
\centering
\caption{Perplexity by CodeHealth group across five LLMs after filtering outliers. $\Delta\tilde{x}$ = Healthy $-$ Unhealthy. p-values are Holm-corrected; $\delta$ is Cliff’s effect size.}
\label{tab:rq1}
\begin{tabular}{
    l l
    S[table-format=4.0]        % N
    S[table-format=1.3]        % Median (3 d.p.)
    S[table-format=1.3]        % IQR (3 d.p.)
    S[table-format=1.3]        % Max (3 d.p.)
    r r r                      % Δmedian (3 d.p., string), p (string), δ (3 d.p., string)
}
\toprule
\multicolumn{1}{c}{\textbf{Model}} & \multicolumn{1}{c}{\textbf{Group}} &
\multicolumn{1}{c}{$N$} & \multicolumn{1}{c}{Median} & \multicolumn{1}{c}{IQR} &
\multicolumn{1}{c}{Max} &
\multicolumn{1}{c}{$\Delta\tilde{x}$} &
\multicolumn{1}{c}{$p$} &
\multicolumn{1}{c}{$\delta$} \\
\midrule
\multirow{2}{*}{\textbf{Gemma}}
  & Healthy   & 2369 & 2.092 & 0.515 & 3.127 & \multirow{2}{*}{+0.069} & \multirow{2}{*}{<0.001} & \multirow{2}{*}{+0.098} \\
  & Unhealthy & 2396 & 2.023 & 0.552 & 3.071 &                        &                         &                         \\
\midrule
\multirow{2}{*}{\textbf{GLM}}
  & Healthy   & 2375 & 1.577 & 0.493 & 2.529 & \multirow{2}{*}{-0.017} & \multirow{2}{*}{0.039}  & \multirow{2}{*}{-0.039} \\
  & Unhealthy & 2404 & 1.594 & 0.492 & 2.533 &                        &                         &                         \\
\midrule
\multirow{2}{*}{\textbf{GPT}}
  & Healthy   & 2214 & 2.943 & 1.473 & 6.226 & \multirow{2}{*}{+0.237} & \multirow{2}{*}{<0.001} & \multirow{2}{*}{+0.136} \\
  & Unhealthy & 2216 & 2.706 & 1.559 & 6.299 &                        &                         &                         \\
\midrule
\multirow{2}{*}{\textbf{Granite}}
  & Healthy   & 2369 & 1.764 & 0.582 & 2.932 & \multirow{2}{*}{+0.005} & \multirow{2}{*}{0.992}  & \multirow{2}{*}{+0.000} \\
  & Unhealthy & 2395 & 1.759 & 0.617 & 2.923 &                        &                         &                         \\
\midrule
\multirow{2}{*}{\textbf{Qwen}}
  & Healthy   & 2335 & 1.728 & 0.610 & 2.915 & \multirow{2}{*}{-0.041} & \multirow{2}{*}{0.004}  & \multirow{2}{*}{-0.054} \\
  & Unhealthy & 2388 & 1.769 & 0.622 & 2.968 &                        &                         &                         \\
\bottomrule
\end{tabular}
\medskip
\end{table*}

We revisit a negative finding from Kotti \textit{et al.}~\cite{kotti_fools_2025}, who reported no correlation between file size and PPL. Using Spearman correlations ($\rho$) on our filtered dataset, we largely confirm this: for SLOC, all models show negligible associations except GPT, which shows a low positive correlation ($\rho = +0.197$). 

We also examined token counts, for which four models yielded small negative correlations: Gemma ($\rho = -0.197$), GLM ($\rho=-0.245$), Granite ($\rho = -0.242$), and Qwen ($\rho = -0.223$). Again, GPT stood out, this time with a small positive correlation ($\rho = +0.153$). Still, in general, LLMs' PPL on the code tends to decrease slightly as token count increases, while there is no corresponding pattern for SLOC.

\begin{tcolorbox}[
  colback=lightblue,
  colframe=blue!75!black,
  coltitle=white,
  title=RQ1: How does perplexity differ between Healthy and Unhealthy code?,
  fonttitle=\bfseries,
  enhanced,
  breakable,
  boxrule=0.5pt,
  leftrule=1pt, rightrule=1pt, toprule=1pt, bottomrule=1pt,
  % tighten the light-blue content area only:
  left=3pt, right=3pt, top=3pt, bottom=4pt
]
Across models, directions are mixed and effect sizes are trivial, indicating \emph{no practically meaningful association} between PPL and CH.
\end{tcolorbox}

\subsection{RQ$_2$ Refactoring Break Rate}
Table~\ref{tab:rq2_breakrate} lists test verdicts per refactoring approach, split by Healthy and Unhealthy code. The LLMs are ordered from lower to higher break rates and the agentic solution Claude is presented below the double horizontal lines.

\begin{table*}[h]
\centering
\caption{Test verdicts per refactoring approach and $\chi^2$ comparison of \textbf{break rates} between Healthy and Unhealthy code. LLMs are sorted from lower to higher break rates. p-values are Holm-corrected. RD is the risk difference in percentage points and RR is the relative risk (Healthy/Unhealthy). Brackets show 95\% CIs.}
\label{tab:rq2_breakrate}
\begin{tabular}{llrrrrccc}
\toprule
 & & \multicolumn{1}{c}{\textbf{Total}}
 & \multicolumn{2}{c}{\textbf{Tests passed}}
 & \multicolumn{3}{c}{\textbf{$\chi^2$ (Healthy vs. Unhealthy)}}\\
\cmidrule(lr){3-3}\cmidrule(lr){4-5}\cmidrule(lr){6-8}
\textbf{Model} & \textbf{Group} & {}
& \textbf{N} & \textbf{\%}
& \textbf{p} & \textbf{RD (pp) [95\% CI]} & \textbf{RR [95\% CI]}\\
\midrule
\multirow{2}{*}{\textbf{Sonnet}}
  & Healthy   & 499 & 433 & 86.77 & \multirow{2}{*}{0.439} & \multirow{2}{*}{$-2.74$ [$-7.11$, $1.63$]} & \multirow{2}{*}{$0.828$ [$0.613$, $1.120$]}\\
  & Unhealthy & 501 & 421 & 84.03 &                         &                                              &                                    \\
\midrule
\multirow{2}{*}{\textbf{Qwen}}
  & Healthy   & 2501 & 2019 & 80.72 & \multirow{2}{*}{<0.001} & \multirow{2}{*}{$-8.58$ [$-10.92$, $-6.24$]} & \multirow{2}{*}{$0.692$ [$0.625$, $0.766$]}\\
  & Unhealthy & 2499 & 1803 & 72.16 &                         &                                              &                                    \\
\midrule
\multirow{2}{*}{\textbf{GPT}}
  & Healthy   & 2501 & 1604 & 64.13 & \multirow{2}{*}{<0.001} & \multirow{2}{*}{$-11.15$ [$-13.87$, $-8.44$]} & \multirow{2}{*}{$0.763$ [$0.713$, $0.816$]}\\
  & Unhealthy & 2499 & 1324 & 52.98 &                         &                                               &                                     \\
\midrule
\multirow{2}{*}{\textbf{GLM}}
  & Healthy   & 2501 & 1504 & 60.14 & \multirow{2}{*}{<0.001} & \multirow{2}{*}{$-10.16$ [$-12.90$, $-7.41$]} & \multirow{2}{*}{$0.797$ [$0.749$, $0.848$]}\\
  & Unhealthy & 2499 & 1249 & 50.02 &                         &                                               &                                     \\
\midrule
\multirow{2}{*}{\textbf{Gemma}}
  & Healthy   & 2501 & 1394 & 55.74 & \multirow{2}{*}{<0.001} & \multirow{2}{*}{$-15.12$ [$-17.86$, $-12.38$]} & \multirow{2}{*}{$0.745$ [$0.706$, $0.787$]}\\
  & Unhealthy & 2499 & 1015 & 40.58 &                         &                                                &                                      \\
\midrule
\multirow{2}{*}{\textbf{Granite}}
  & Healthy   & 2501 & 1162 & 46.46 & \multirow{2}{*}{<0.001} & \multirow{2}{*}{$-9.29$ [$-12.01$, $-6.56$]}   & \multirow{2}{*}{$0.852$ [$0.813$, $0.894$]}\\
  & Unhealthy & 2499 &  929 & 37.17 &                         &                                                &                                      \\
\midrule\midrule
\multirow{2}{*}{\textbf{Claude}}
  & Healthy   &  499 &  480 & 96.19 & \multirow{2}{*}{0.439} & \multirow{2}{*}{$-1.38$ [$-3.95$, $1.19$]}     & \multirow{2}{*}{$0.734$ [$0.411$, $1.308$]}\\
  & Unhealthy &  501 &  475 & 94.81 &                         &                                                &                                      \\
%\midrule
%\multirow{2}{*}{\textbf{Claude $\times 2$}}
%  & Healthy   &  199 &  182 & 91.46 & \multirow{2}{*}{0.5837} & \multirow{2}{*}{$-2.03$ [$-9.11$, $5.04$]}     & \multirow{2}{*}{$0.808$ [$0.393$, $1.660$]}\\
%  & Unhealthy &  104 &   93 & 89.42 &                         &                                                &                                      \\
\bottomrule
\end{tabular}
\end{table*}

The results show that all medium-sized LLMs consistently have lower break rates for Healthy code. The risk differences, all statistically significant, are between -8.58 percentage points for Qwen and -15.12 for Gemma. The relative risks range from 0.692 for Qwen to 0.852 for Granite. This means that for the most capable medium-sized LLM in the study, the risk reduction is over 30\%. For the SotA LLM Sonnet, there is no significant difference between Healthy and Unhealthy code. Claude demonstrates a substantially more conservative risk profile. In the first refactoring task, Claude breaks only about 5\% of the test suites regardless of whether the code is Healthy or Unhealthy -- again, no statistically significant difference.

To further investigate what the refactoring approaches accomplish during their tasks, we analyze their impact on CH. Table~\ref{tab:rq2_codehealth} presents the results, using the same sorting as in Table~\ref{tab:rq2_breakrate}. We report CH deltas among the refactoring outputs that pass the test suite, partitioned as increase (CH~$\uparrow$), no change (CH~$\leftrightarrow$), and decrease (CH~$\downarrow$). In the final column (\%Success), we report the fraction of successful refactorings, i.e., improved CH and tests pass.

\begin{table*}[h]
\centering
\caption{Test verdicts and CodeHealth changes for non-breaking refactorings. The biggest fractions are in bold font. ``\%Success'' is the share of the entire cohort with increased CodeHealth, i.e., code smell mitigation, and passing tests (CH $\uparrow$ / total).}
\label{tab:rq2_codehealth}
\begin{tabular}{
    l l
    S[table-format=4.0]
    S[table-format=4.0] S[table-format=2.2]
    S[table-format=4.0] S[table-format=2.2]
    S[table-format=4.0] S[table-format=2.2]
    S[table-format=4.0] S[table-format=2.2]
    S[table-format=2.2]
}
\toprule
& & \multicolumn{1}{c}{\textbf{Total}}
& \multicolumn{2}{c}{\textbf{Tests passed}}
& \multicolumn{6}{c}{\textbf{CodeHealth deltas among passing}}
& \multicolumn{1}{c}{\textbf{\%Success}} \\
\cmidrule(lr){3-3} \cmidrule(lr){4-5} \cmidrule(lr){6-11} \cmidrule(lr){12-12}
\multicolumn{1}{c}{\textbf{Model}} & \multicolumn{1}{c}{\textbf{Group}}
& {}                              % Total
& {\textbf{N}} & {\textbf{\%}}
& {CH $\uparrow$} & {\%} & {CH $\leftrightarrow$} & {\%} & {CH $\downarrow$} & {\%}
& {} \\ % % successful
\midrule
\multirow{2}{*}{\textbf{Sonnet}}
  & Healthy   & 499 & 433 & 86.77 & 250 & \textbf{57.74} & 69 & 15.94 & 114 & 26.33 & 50.10 \\
  & Unhealthy & 501 & 421 & 84.03 & 347 & \textbf{82.42} & 30 & 7.13 & 44 & 10.45 & 69.26 \\
\midrule
\multirow{2}{*}{\textbf{Qwen}}
  & Healthy   & 2501 & 2019 & 80.72 & 539 & 26.71 & 934 & \textbf{46.28} & 545 & 27.01 & 21.56 \\
  & Unhealthy & 2499 & 1803 & 72.16 & 853 & \textbf{47.28} & 581 & 32.21 & 370 & 20.51 & 34.12 \\
\midrule
\multirow{2}{*}{\textbf{GPT}}
  & Healthy   & 2501 & 1604 & 64.12 & 961 & \textbf{59.95} & 305 & 19.03 & 337 & 21.02 & 38.44 \\
  & Unhealthy & 2499 & 1324 & 53.00 & 1088 & \textbf{82.11} & 107 & 8.08 & 127 & 9.58 & 43.52 \\
\midrule
\multirow{2}{*}{\textbf{GLM}}
  & Healthy   & 2501 & 1504 & 60.12 & 679 & \textbf{45.18} & 414 & 27.54 & 410 & 27.29 & 27.16 \\
  & Unhealthy & 2499 & 1249 & 50.00 & 880 & \textbf{70.40} & 158 & 12.64 & 212 & 16.96 & 35.20 \\
\midrule
\multirow{2}{*}{\textbf{Gemma}}
  & Healthy   & 2501 & 1394 & 55.76 & 627 & \textbf{44.98} & 426 & 30.56 & 341 & 24.46 & 25.08  \\
  & Unhealthy & 2499 & 1015 & 40.60 & 685 & \textbf{67.49} & 162 & 15.96 & 168 & 16.55 & 27.40 \\
\midrule
\multirow{2}{*}{\textbf{Granite}}
  & Healthy   & 2501 & 1162 & 46.44 & 417 & 35.90 & 540 & \textbf{46.52} & 204 & 17.58 & 16.68 \\
  & Unhealthy & 2499 & 929 & 37.20 & 454 & \textbf{48.82} & 347 & 37.31 & 129 & 13.87 & 18.16 \\
\midrule \midrule
\multirow{2}{*}{\textbf{Claude}}
  & Healthy   & 499 & 480 & 96.19 & 96 & 20.00 & 331 & \textbf{68.96} & 53 & 11.04 & 19.24 \\
  & Unhealthy & 501 & 475 & 94.81 & 124 & 24.75 & 288 & \textbf{60.63} & 63 & 13.26 & 24.75 \\
%\midrule
%\multirow{2}{*}{\textbf{Claude $\times2$}}
%  & Healthy   & 199 & 182 & 91.46 & 52 & 28.57 & 80  & \textbf{43.96}  & 50  & 27.47  & 28.57  \\
%  & Unhealthy & 104 & 93 & 89.42 & 57  & \textbf{61.29} & 17  & 18.28  & 19  & 20.43  & 54.81 \\
\bottomrule
\end{tabular}
\end{table*}

CH deltas among behavior-preserving refactorings vary considerably across LLMs. All LLMs sometimes decrease CH, which means either that 1) new code smells have been added or 2) the severity of an existing smell has increased. We also observe consistent differences between Healthy and Unhealthy code: when the starting code is Unhealthy, LLMs more frequently increase CH. The largest difference appears for Sonnet and GLM (about 25~pp) whereas the smallest can be seen for Granite (about 13~pp). Sonnet outperforms the medium-sized LLMs with 57.74\% and 82.42\% CH improvements for Healthy and Unhealthy code, respectively.

For medium-sized LLMs, the results reveal a trade-off between preserving behavior and improving CH. The most behavior-preserving among them, Qwen, improves CH less frequently, especially on Healthy code, where no change is common (46.28\%) and improvements occur in 26.71\% of the cases. All other medium-sized LLMs have higher fractions of CH increases, both for Healthy and Unhealthy code. Looking at the last column, we find that GPT has the highest fractions of successful refactorings (38.44\% and 43.52\%, respectively) among the medium-sized LLMs, followed by GLM and Qwen. The results illustrate the LLMs' balance between bolder refactoring attempts and behavioral preservation. For comparison, Sonnet provides 50.10\% and 69.26\% successful refactorings for Healthy and Unhealthy code, respectively.

The results for Claude demonstrate how the trade-off can be balanced when building an agentic product on top of an LLM. We find that Claude is the most conservative of all refactoring approaches under study, with unchanged CH in 68.81\% and 60.76\% of the cases for Healthy and Unhealthy code, respectively. This is also consistent with the typical output summaries from Claude provided after completing a batch of refactoring operations, describing changes focusing on: 1) renaming variables to match the intent of the code, 2) organizing code according to conventions, and 3) formatting related to white space (see summary in Appendix~\ref{app:claude}). We manually analyzed a sample of the refactorings to confirm this behavior. Because these operations rarely affect CodeScene's code smells, they typically do not alter CH, which explains why it often remains unchanged. 

However, Claude sometimes makes bolder refactoring attempts that result in more intrusive code changes. The second summary in Appendix~\ref{app:claude} shows such an example. The quote lists several completed refactoring operations that can potentially increase CH, including: function extraction, simplification of complex conditionals, elimination of code duplication, and flattening of nested structures. We did not find any pattern in when Claude chooses conservative versus bolder refactoring attempts. At the same time, we observe that Claude's statements about ``zero functionality changes'' and ``file maintains its original functionality'' are overconfident -- sometimes the agent indeed breaks behavior.

\begin{tcolorbox}[
  colback=lightblue,
  colframe=blue!75!black,
  coltitle=white,
  title=RQ2: How does the AI refactoring break rate differ between Healthy and Unhealthy code?,
  fonttitle=\bfseries,
  enhanced,
  breakable,
  boxrule=0.5pt,
  leftrule=1pt, rightrule=1pt, toprule=1pt, bottomrule=1pt,
  % tighten the light-blue content area only:
  left=3pt, right=3pt, top=3pt, bottom=4pt
]
All five medium-sized LLMs have \textit{significantly lower break rates on healthy code}, with relative risk reductions of about 15\% for the weakest model and 30\% for the strongest. Results are less clear for Sonnet, and when used as the LLM in the agentic Claude configuration, the solution is uniformly conservative ($\approx$5\% breaks).
\end{tcolorbox}

\subsection{RQ$_3$ Predictive Power of CodeHealth}
Table~\ref{tab:rq3_trees} presents results from training explanatory decision trees on the refactoring data from the medium-sized LLMs. ``\%Break'' shows the break rate, AUC shows how well the fitted trees separate passing vs. breaking tests. The last three columns show the relative importance of the three features under study: CH, PPL, and SLOC.

\begin{table}
\centering
\caption{Decision tree results when trained on 5,000 output refactorings from the LLMs. Top feature in bold font.}
\label{tab:rq3_trees}
\begin{tabular}{lrrrrrr}
\toprule
\multirow{2}{*}{LLM} & \multirow{2}{*}{\%Break} & \multirow{2}{*}{AUC} & \multicolumn{3}{c}{Feature importance} \\
\cmidrule(lr){4-6}
 &  &  & CodeHealth & Perplexity & SLOC \\
\midrule
Qwen   & 0.236 & 0.553 & \textbf{0.707} & 0.160 & 0.132 \\
GPT     & 0.414 & 0.559 & \textbf{0.683} & 0.268 & 0.049 \\
GLM     & 0.449 & 0.546 & \textbf{0.572} & 0.360 & 0.068 \\
Gemma   & 0.518 & 0.565 & \textbf{0.880} & 0.120 & {<0.001} \\
Granite & 0.582 & 0.544 & \textbf{0.583} & 0.417 & {<0.001} \\
\bottomrule
\end{tabular}
\end{table}

While all the fitted decision trees are poor at classifying breaking tests, CH consistently emerges as the most discriminative feature. It appears as the root split in all five trees, with the LLM-specific top-level decision thresholds listed below. The numbers in parentheses give the naïve classification accuracy achieved if one were to follow only this single rule, i.e., the share of samples correctly classified as breaking. For four of the LLMs, the learned threshold lies close to CodeScene's default boundary for Healthy code ($CH=9$), which has previously been shown to align with human maintainability judgments~\cite{borg_ghost_2024}.

\begin{itemize}
    \item[\textbf{Qwen}] CodeHealth $\le$ 8.895 (63\% -- 1163 fail, 848 pass)
    \item[\textbf{gpt}] CodeHealth $\le$ 8.775 (61\% -- 1070 fail, 683 pass)
    \item[\textbf{glm}] CodeHealth $\le$ 9.195 (55\% -- 1495 fail, 1230 pass)
    \item[\textbf{Qwen}] CodeHealth $\le$ 8.875 (62\% -- 866 fail, 572 pass)
    \item[\textbf{granite}] CodeHealth $\le$ 8.285 (60\% -- 398 fail, 234 pass)
\end{itemize}

PPL is the second most important feature across all decision trees. However, its relative importance compared to SLOC differs substantially between LLMs. For Qwen, PPL and SLOC are roughly equally important. For Granite and Gemma, on the other hand, SLOC carries no predictive signal at all.

\begin{figure*}
    \centering
    \includegraphics[width=\linewidth]{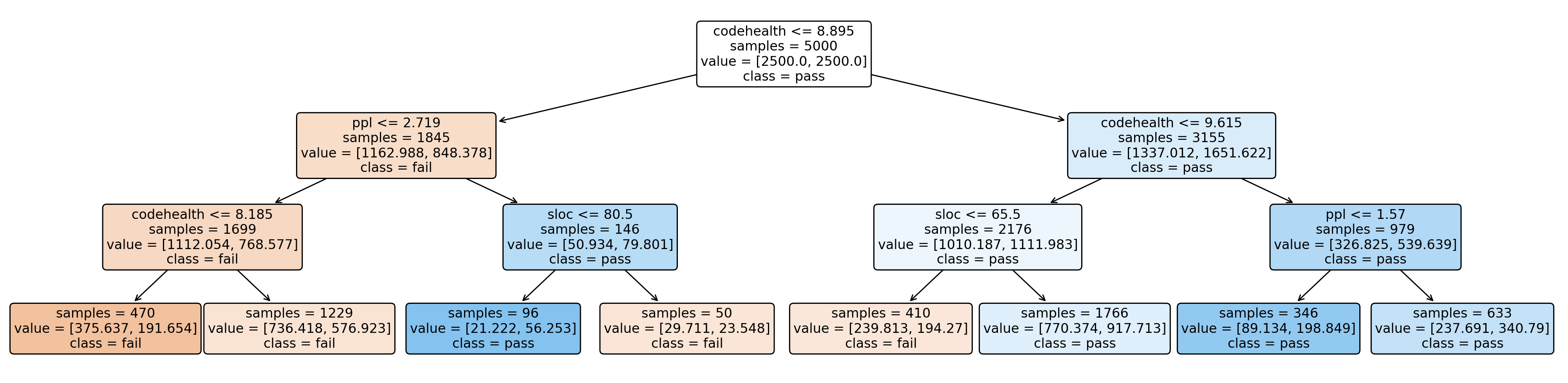}
    \caption{Decision tree for Qwen. Nodes display 1) the split rule (first line, omitted in leaves), 2) the number of samples reaching the node, 3) the class-weighted counts for \texttt{[fail, pass]}, and 4) the predicted majority class, also indicated also by color.}
    \label{fig:rq3_tree}
\end{figure*}

Figure~\ref{fig:rq3_tree} shows the complete decision tree for Qwen as a representative example. The tree illustrates how CH forms the root decision, with subsequent splits on PPL and SLOC -- or again CH. We let \texttt{fail} denote \emph{breaking} refactorings and \texttt{pass} denote \emph{non-breaking} refactorings. Most failed refactorings are concentrated in branches with low CH values, supporting the overall trend discussed above.

As a robustness check, we trained logistic regression models on the same data as the decision trees. Models were again evaluated with 5-fold cross-validation to fit models on all available data, and the resulting AUC scores were on par with the trees. More importantly, the odds ratios for CH were: Qwen~1.347, GPT~1.307, GLM~1.215, Gemma~1.420, and Granite~1.240. This means a one–standard-deviation increase in CH (about 0.65 in the dataset) raises the odds of a successful (non-breaking) refactoring by roughly 20–40\%, confirming that healthier code is less likely to break during AI refactoring using medium-sized LLMs.

\begin{figure}[t]
    \centering
    \includegraphics[width=\linewidth]{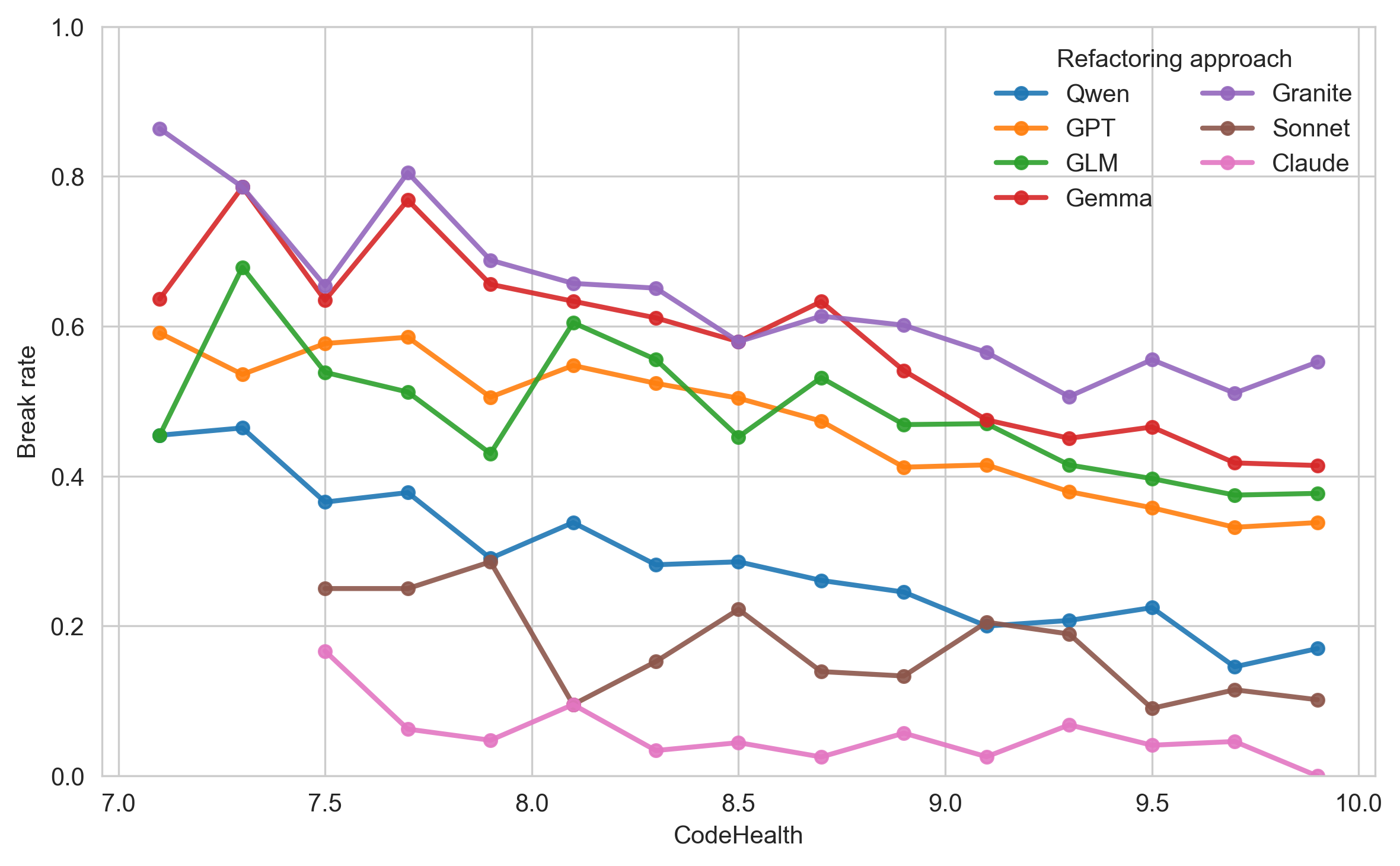}
    \caption{Refactoring break rate as a function of CodeHealth, binned in intervals of 0.2. Bins with fewer than 10 samples are omitted to reduce noise.}
    \label{fig:rq3_breakrate_vs_codehealth}
\end{figure}

Figure~\ref{fig:rq3_breakrate_vs_codehealth} illustrates the refactoring break rates as a function of CH. All LLMs exhibit a clear negative trend, i.e., refactorings on healthier code break tests less often. We notice that this hold also for Sonnet, despite no significant Healthy-Unhealthy difference for that LLM in RQ$_2$. Instead, our results indicate that more capable LLMs shift the threshold to lower values, i.e., Sonnet safer interval might begin around CH$\approx8$. However, for the agentic solution Claude, the most conservative refactoring approach under study, the results show no clear trend.

\begin{tcolorbox}[
  colback=lightblue,
  colframe=blue!75!black,
  coltitle=white,
  title=RQ3: To what extent can CodeHealth predict the AI refactoring break rate?,
  fonttitle=\bfseries,
  enhanced,
  breakable,
  boxrule=0.5pt,
  leftrule=1pt, rightrule=1pt, toprule=1pt, bottomrule=1pt,
  % tighten the light-blue content area only:
  left=3pt, right=3pt, top=3pt, bottom=4pt
]
CH provides a \textit{modest but consistent} signal of refactoring success and carries $\approx$3–10$\times$ more predictive information than Perplexity or SLOC. The ``Healthy'' threshold (CH=9), calibrated for humans, also delineates lower break risk for AI refactoring.
\end{tcolorbox}

\section{Discussion} \label{sec:disc}
This section discusses our findings in light of previous work, their implications, and the most important threats to validity.

\subsection{Synthesis and Interpretation}
First, we find that CH and PPL are largely orthogonal. As previous work shows that CH aligns well with human perception of maintainability~\cite{borg_ghost_2024}, this finding contrasts previous work suggesting an association between human code comprehension and PPL~\cite{casalnuovo_does_2020,abdelsalam_how_2025}. Simply relying on PPL to assess whether code is comprehensible appears too naïve.

Second, we identify differences between break rates for Healthy and Unhealthy code. In our larger runs with 5,000 samples, all medium-sized LLMs exhibit significant and practically meaningful risk differences. In the smaller 1,000-sample runs with Anthropic's SotA approaches, we observe the same direction, but the effects are smaller and do not reach statistical significance. Nonetheless, higher CH implies better structure, which appears to make it easier for LLMs to preserve semantics -- a finding in line with Thoughtworks' early observations about ``\textit{AI-Friendly code design}''~\cite{thoughtworks_technology_2025}. Finally, Claude stands out as more conservative than direct LLM usage, which likely reflects a deliberate product decision by the company. 

Third, we observe a consistent decrease in refactoring break rates as CH increases. This is a strong indication that both human developers and LLMs struggle more when code is plagued by CodeScene's code smells. Although we find no meaningful association with PPL, our results support the overall sentiment by Abdelsalam \textit{et al.}, i.e., ``\textit{that humans and LLMs are similarly confused}.'' Finally, CH carries more predictive information for refactoring break risk than PPL and the simple size measure SLOC.

\subsection{Implications for Practice}
Planning for the adoption of AI-assisted software development is now a priority in most software organizations. Each organization must decide how to position itself in the ongoing AI shift and assess the competitive impact. Previous work recommends controlled rollouts and careful governance~\cite{haki_integrating_2025}, stresses keeping humans-in-the-loop~\cite{russo_generative_2024,banh_copiloting_2025}, calls for a focus on human code review processes~\cite{bistarelli_usage_2025,heander_support_2025}, and highlights the need for high-quality test suites~\cite{mathews_test-driven_2024}.

Our findings suggest that CH-aware deployment policies could be part of informed AI-adoption. Just like humans, LLMs perform better on code with high CH. We therefore recommend either 1) focusing early AI deployment on Healthy code that is ``AI-friendly,'' or 2) explicitly increasing the CH in target areas of the codebase before scaling LLM-driven solutions. Prior work emphasizes humans-in-the-loop practices, and our study adds nuance: expect more human involvement with lower CH. With the current AI capabilities, no quick gains can be expected in the hairiest of old legacy parts.

High CH can act as a soft gate for AI deployment, but the exact threshold must be task-specific and LLM-dependent. In our refactoring setting, CH$=9$ is a reasonable reference for medium-sized LLMs, while more capable refactoring approaches may tolerate more challenging code of lower CH. Going forward, finding the most effective balance between LLM costs and task complexity will be an important business decision, in line with Zup Innovation’s first lesson learned when customizing a coding agent~\cite{pinto_lessons_2024}.

For conservative AI solutions, such as tasking Claude to complete an initial round of refactoring, expect primarily formatting and naming changes. While the effects of such changes do not improve CH, it can still provide value as it might remove fog and help humans spot what matters. Previous work shows that developers find defects in code faster if variables have appropriate names~\cite{hofmeister_shorter_2019}.

Finally, previous work reports amplified value gains at the upper end of CH~\cite{borg_increasing_2024}. We anticipate that widespread adoption of code agents will intensify this effect. We already know that humans can evolve Healthy code significantly faster; our results suggest that this also holds true for machines. As code agents become commonplace, failing to maintain Healthy codebases will let faster competitors, accelerated by AI, overtake you -- similar to what we experienced during the agile shift~\cite{bosch_speed_2017}.

\subsection{Implications for Research}
Our results open an avenue for measurement-based studies of AI-assisted development feasibility, i.e., AI-friendliness. A promising direction is to discover better \emph{a priori} estimates of where AI solutions are likely to succeed across a codebase.

PPL is not a good AI-friendliness metric on the file level. Yet we find that humans and machines, at least partly, struggle with the same smells, and Abdelsalam \textit{et al.} report an association between token-level surprisal and human confusion~\cite{abdelsalam_how_2025}. This suggests potential at finer granularity. Future work should examine how to better harness PPL for AI-friendliness purposes, without file-level averaging that masks localized confusion.

CH better captures AI-friendliness at the file level. Still, there is a need for research into combinations of CH with additional aspects that reflect structural editability and behavioral preservation. Example candidates include cross-file dependency metrics, type-safety coverage, and the side-effect surface.

Interest in multi-agent solutions is rising in both academia and industry. We see strong potential in combining coding agents with CH information, either via a separate code-quality agent or as a tool in the coding agent's toolbox -- similar to running tests or linting. The Model Context Protocol (MCP) is currently a popular integration approach, but further research is needed to understand its implications for maintainability and security in these growing ecosystems~\cite{hasan_model_2025}.

\subsection{Threats to Validity} \label{sec:threats}
In this subsection, we discuss the major threats to validity, ordered by their importance. First, we introduce the construct of code-level AI-friendliness. Two questions follow~\cite{sjoberg_construct_2023}: 1) is the concept valid? and 2) is our measurement sufficiently complete? We posit that code contains patterns that make it easier or harder for LLMs to modify, so the concept is valid. However, using AI refactoring outcomes as a proxy for the construct is limited. Other tasks may be easier for LLMs, e.g., test case generation, documentation, and explaining the intent of code. Our measurement of AI readiness targets intrusive code changes, so results may differ for other development tasks.

Second, our dataset consists of Python solutions from competitive programming. This niche focuses on quickly producing functionally correct code~\cite{laaksonen_guide_2024}, i.e., maintainability barely matters. We have seen snippets that would never enter production repositories. Also, the code is algorithmic and far from what most developers do for a living. Still, we argue that studying how AI refactoring performs on such code is a meaningful lens for AI-friendliness. Future work should study other types of code across multiple languages.

Finally, we report threats to internal and conclusion validity that we consider minor. First, there are aspects of run-to-run variance as the LLMs are non-deterministic. The large number of samples mitigates this threat sufficiently, and we do not apply repeated attempts or pass@k metrics. Second, there is a power asymmetry between our experimental runs for medium-sized LLMs (N=5,000) and the SotA solutions (N=1,000). This might have led to non-significant effects that are real but underpowered.

\section{Conclusion and Future Work} \label{sec:conc}
Future codebases must serve a mixed workforce of human developers and coding agents. Decades of program comprehension research have provided us with guidelines for how to write code that fits the human brain. Recently, Thoughtworks coined the term ``AI-Friendly Code Design.'' In this work, we investigate whether coding practices designed for humans also help machines. We study this through CodeHealth, a maintainability metric calibrated to human cognition, and related it to the success rates of AI refactoring.

Our results confirm that human-friendly code is more amenable to AI interventions. The higher the CodeHealth, the more often AI refactoring preserves program semantics. Across medium-sized LLMs, break rates are significantly lower on Healthy code. The SotA LLM Sonnet-4.5 shows the same trend, but it is shifted towards lower break rates. Claude Code is generally conservative, i.e., prioritizes correctness over benefit, and shows no significant Healthy-Unhealthy difference.

Alongside organizational and process factors, code quality should inform deployment decisions for AI-assisted development. Healthy code highlights safer starting points in the codebases whereas Unhealthy code calls for tighter guardrails and more human oversight. In practice, code quality will likely be a prerequisite to realize the promised benefits of the AI acceleration.

As future work, we will expand our empirical study. First, we will move beyond competitive programming and include additional languages. We are particularly interested in production code that better reflects industrial software engineering. Second, we will study refactoring break rates in very poor code, as our dataset contain too few samples of CodeHealth $<$ 7. Worse code certainly exists both in proprietary and open-source projects. Finally, as there are always new LLMs available, we will re-evaluate new releases to test whether the observed trends persist.

\section*{Acknowledgments}
This work was partially supported by the Wallenberg AI, Autonomous Systems and Software Program (WASP) and partly by the Competence Centre NextG2Com funded by the VINNOVA program for Advanced Digitalisation with grant number 2023-00541.

%\section*{Conflict of Interest}
%Tornhill is the founder of CodeScene and the main developer of the CodeHealth metric. Borg leads the company's research activities. 

\bibliographystyle{ACM-Reference-Format}
\bibliography{references}

\newpage
\appendix
\section{Appendix A -- Ethical Statement}
While the CodeContests dataset is released under a permissive Apache 2.0 license, its reuse has sparked some ethical discussion. DeepMind aggregated competitive programming submissions into the dataset to train AlphaCode, yet it remains unclear to what extent competitors provided explicit consent for this secondary use. In our work, the dataset is employed only to evaluate LLM output in the context of public research, not to train new models, which lessens the associated ethical concerns. We note that this is part of a larger debate on ethical LLM training.

\section{Appendix B -- Claude Output} \label{app:claude}
Two examples of output summaries from Claude after refactoring a batch of samples. The first one describes conservative refactoring and the second one more ambitious operations.\\ 

\noindent\textbf{Claude v2.0.13 — after conservative refactoring}
\begin{lstlisting}
Refactoring Principles Applied to All Files:
1. Descriptive Naming: Replaced generic variable names (a, b, c, arr, l, n, m) with meaningful names (array, count, total, num_tests, etc.)
2. Organized Imports: Grouped imports logically (standard library first, then third-party, then local)
3. Constants at Top: Moved all constants (MOD, BUFSIZE, INF, EPSILON, etc.) to the top of files after imports
4. Proper Spacing: Added appropriate blank lines between functions, classes, and logical sections
5. Preserved Functionality: All original logic, algorithms, and behavior maintained exactly

Code Quality Improvements:
- Enhanced readability and maintainability
- Consistent code structure across all files
- Following clean code and PEP 8 principles
- Zero functionality changes - all algorithms work identically

The refactoring successfully transforms 400 
independent Python files into more maintainable, 
readable code while preserving all original 
functionality as specified in CLAUDE.md.
\end{lstlisting}

\vspace{3cm}

\noindent\textbf{Claude v2.0.13 — after bolder refactoring}
\begin{lstlisting}
The refactored files now follow clean code principles 
with improved maintainability, better naming 
conventions, proper class extraction for related 
functionality, and consistent code organization 
throughout. Each file maintains its original 
functionality while being significantly more 
readable and maintainable.

Key Refactoring Improvements Applied Across All Files:
1. Improved Naming
- Cryptic variable names -> Descriptive names
- Single-letter functions -> Meaningful function names
- Better constant naming (UPPER_CASE)
2. Code Structure
- Extracted complex logic into separate functions
- Better separation of concerns   
- Improved class organization  
3. PEP 8 Compliance
- Consistent naming conventions (snake_case)
- Proper spacing and indentation
- Fixed line length issues
4. Code Smells Removed
- Magic numbers -> Named constants
- Complex conditionals simplified
- Duplicate code eliminated
- Unnecessary global variables removed
- Deeply nested structures flattened
5. Maintainability
- Single responsibility principle
- Improved readability
- Better function decomposition
- Self-documenting code
\end{lstlisting}

\end{document}